\documentclass[preprint,aps,amsmath,showpacs,showkeys,nofootinbib, 
superscriptaddress]{revtex4} 
\usepackage[dvips]{graphicx}
\usepackage{color}  

\begin{document} 

\title{$\Lambda\Lambda N - \Xi NN$ $S$ wave resonance}
\author{H. Garcilazo} 
\email{humberto@esfm.ipn.mx} 
\affiliation{Escuela Superior de F\' \i sica y Matem\'aticas, \\ 
Instituto Polit\'ecnico Nacional, Edificio 9, 
07738 M\'exico D.F., Mexico} 
\author{A.~Valcarce} 
\email{valcarce@usal.es} 
\affiliation{Departamento de F\'\i sica Fundamental,\\ 
Universidad de Salamanca, 37008 Salamanca, Spain}
\date{\today} 
\begin{abstract}
We use an existing model of the $\Lambda\Lambda N - \Xi NN$ three-body system based in two-body
separable interactions to study the
$(I,J^P)=(1/2,1/2^+)$ three-body channel. For the
$\Lambda\Lambda$, $\Xi N$, and 
$\Lambda\Lambda - \Xi N$ amplitudes we have constructed separable
potentials based on the most recent results of the HAL QCD Collaboration.
They are characterized by the existence of a
resonance just below or above the $\Xi N$ threshold in the so-called 
$H$-dibaryon channel, $(i,j^p)=(0,0^+)$. 
A three-body resonance appears {2.3} MeV above the $\Xi d$ threshold.
We show that if the $\Lambda\Lambda - \Xi N$ $H$-dibaryon channel is not
considered, the $\Lambda\Lambda N - \Xi NN$ $S$ wave resonance disappears.
Thus, the possible existence of a $\Lambda\Lambda N - \Xi NN$
resonance would be sensitive to the $\Lambda\Lambda - \Xi N$ interaction.
The existence or nonexistence of this resonance could be evidenced by
measuring, for example, the $\Xi d$ cross section.
\end{abstract}
\maketitle 

The $\Lambda\Lambda-\Xi N$ system in a pure $S$ wave configuration has 
quantum numbers $(i,j^p)=(0,0^+)$ so that adding one more nucleon, the
$\Lambda\Lambda N-\Xi NN$ system has necessarily 
quantum numbers $(I,J^P)=(1/2,1/2^+)$. In a series of works based on a 
chiral constituent quark model~\cite{GAR1,GAR2,GAR3} this system was studied 
under the assumption that the $H$ dibaryon~\cite{JAF1} has the lower limit mass
determined by the E373 experiment at KEK~\cite{Tak01} from the observation of
a $^6_{\Lambda\Lambda}$He double hypernucleus. 
Despite the large amount of experimental and theoretical efforts, the existence
of the $H$ dibaryon remains inconclusive, see Ref.~\cite{Fra19} for a recent update.
Experimental evidence disfavors large binding energies~\cite{Nak10}, as predicted in
Ref.~\cite{JAF1}, and the high statistics study of $\Upsilon$ decays at 
Belle~\cite{Kim13} found no indication of an $H$ dibaryon with a mass near 
the $\Lambda\Lambda$ threshold. Recently, the HAL QCD 
Collaboration~\cite{HALQCD} has published
a $N_f = 2 + 1 $ study of coupled channel ($\Lambda\Lambda$ and $\Xi N$)
baryon-baryon interactions with near-physical quark masses, namely
$m_\pi=146$~MeV, concluding that the $H$ dibaryon could be a $\Lambda\Lambda$ 
resonance just below or above the $\Xi N$ threshold. Similar results were
obtained in a low-energy effective field theory study of the
$H$ dibaryon in $\Lambda\Lambda$ scattering~\cite{Yam16}. 

The HAL QCD results are being used as input for the study of
strangeness~$-$~2 baryon-baryon interactions, as recently done in
relativistic chiral effective field theory studies~\cite{Lih18}.
The HAL QCD $\Xi N$ interactions have been also
recently used to study the possible existence of $\Xi NN$ bound 
states in Ref.~\cite{HIYAM}, with negative results for 
the $(I,J^P)=(1/2,1/2^+)$ channel. For the $NN$ interaction
they used the AV8 potential~\cite{AV8}.
As the coupling between $\Lambda\Lambda$ and $\Xi N$ is found to be weak in
Ref.~\cite{HALQCD}, they used an effective single-channel $\Xi N$ potential 
in which the coupling to $\Lambda\Lambda$ in $^{11}S_0$ was renormalized 
into a single Gaussian form chosen
to reproduce the $\Xi N$ phase shift obtained with channel coupling.
The three-body $\Xi NN$ problem is solved in the real axis
by means of a variational method with Gaussian bases, the 
Gaussian Expansion Method~\cite{Hiy1,Hiy2}.
The full coupling between the $\Xi NN$ and $\Lambda\Lambda N$ 
channels was not explicitly considered. A similar calculation
based on the Nijmegen ESC08c potentials~\cite{NAG1,NAG2,NAG3} was presented in 
Ref.~\cite{GARV} also with negative results for 
the $(I,J^P)=(1/2,1/2^+)$ channel, see Fig. 2(a) of Ref.~\cite{GARV}. 

Unlike the calculation in Ref.~\cite{HIYAM}, we developed in Ref.~\cite{GAR4} a model of the
$\Lambda\Lambda N - \Xi NN$ three-body system 
which allowed us to look for possible three-body resonances. 
Using separable two-body interactions fitted to the low-energy data of the Nijmegen 
$S$ wave baryon-baryon amplitudes~\cite{NAG1,NAG2,NAG3},
we found a resonance just below the $\Xi d$ threshold with 
a very small width of only 0.09~MeV~\footnote{It is worth to note that the results
for the $\Xi NN$ system with maximal isospin have been
independently reproduced within the integral Faddeev equation 
formalism~\cite{VLA17} in agreement with high accuracy.}.
Qualitatively similar results have been obtained 
in Ref.~\cite{HIYAM}, although as stressed in this manuscript
they are numerically different due to a different $NN$ potential 
and different treatment of ESC08c Nijmegen $S$ wave baryon-baryon
interactions. Such dependencies on the models and parametrizations
of the two-body interactions make this three-body system ideally suited 
for testing the different models for the two-body interactions. 

However, contrary to the recent results of the HAL QCD Collaboration,
the Nijmegen baryon-baryon interactions gave no indication of either a
bound state or a resonance in the $\Lambda\Lambda - \Xi N$ $(0,0^+)$ 
two-body channel, the $H$-dibaryon channel. 
It is thus interesting to see if the existence of a 
resonance just below or above the $\Xi N$ threshold, as it has been found by the HAL QCD 
Collaboration~\cite{HALQCD} and low-energy effective field theory 
studies~\cite{Yam16}, may affect the position 
of the three-body $S$ wave $(1/2,1/2^+)$ $\Lambda\Lambda N - \Xi NN$ resonance
found in Ref.~\cite{GAR4}. 
For this purpose, we have now constructed separable potential models of 
the $\Lambda\Lambda$, $\Xi N$, and $\Lambda\Lambda-\Xi N$ amplitudes reproducing 
the behavior of the HAL QCD Collaboration results~\cite{HALQCD} and we have
performed a full fledged coupled-channel study 
of the $\Lambda\Lambda N - \Xi NN$ three-body system.

We use rank-one separable potentials for all the uncoupled 
two-body channels, that is, for all channels except the $\Lambda\Lambda-\Xi N$
$(0,0^+)$ interaction. They are as follows,
\begin{equation}
V_i^\rho = g_i^\rho\rangle \lambda\langle g_i^\rho \, ,
\label{eq2}
\end{equation}
such that the two-body t-matrices are 
\begin{equation}
t_i^\rho = g_i^\rho\rangle \tau_i^\rho\langle g_i^\rho \, ,
\label{eq3}
\end{equation}
with
\begin{equation}
\tau_i^\rho = \frac{\lambda}{1-
\lambda\langle g_i^\rho|G_0(i)|g_i^\rho\rangle} \, ,
\label{eq4}
\end{equation}
where $G_0(i)=1/(E-K_i+i\epsilon)$ and
$K_i$ is the kinetic energy operator of channel $i$. 
We use Yamaguchi form factors~\cite{YAMA} for the separable
potentials of Eq.~(\ref{eq2}), i.e., 
\begin{equation}
g_i^\rho(p)=\frac{1}{\alpha^2+p^2}.
\label{eq79}
\end{equation}

\begin{figure*}[t]
\resizebox{8.cm}{12.cm}{\includegraphics{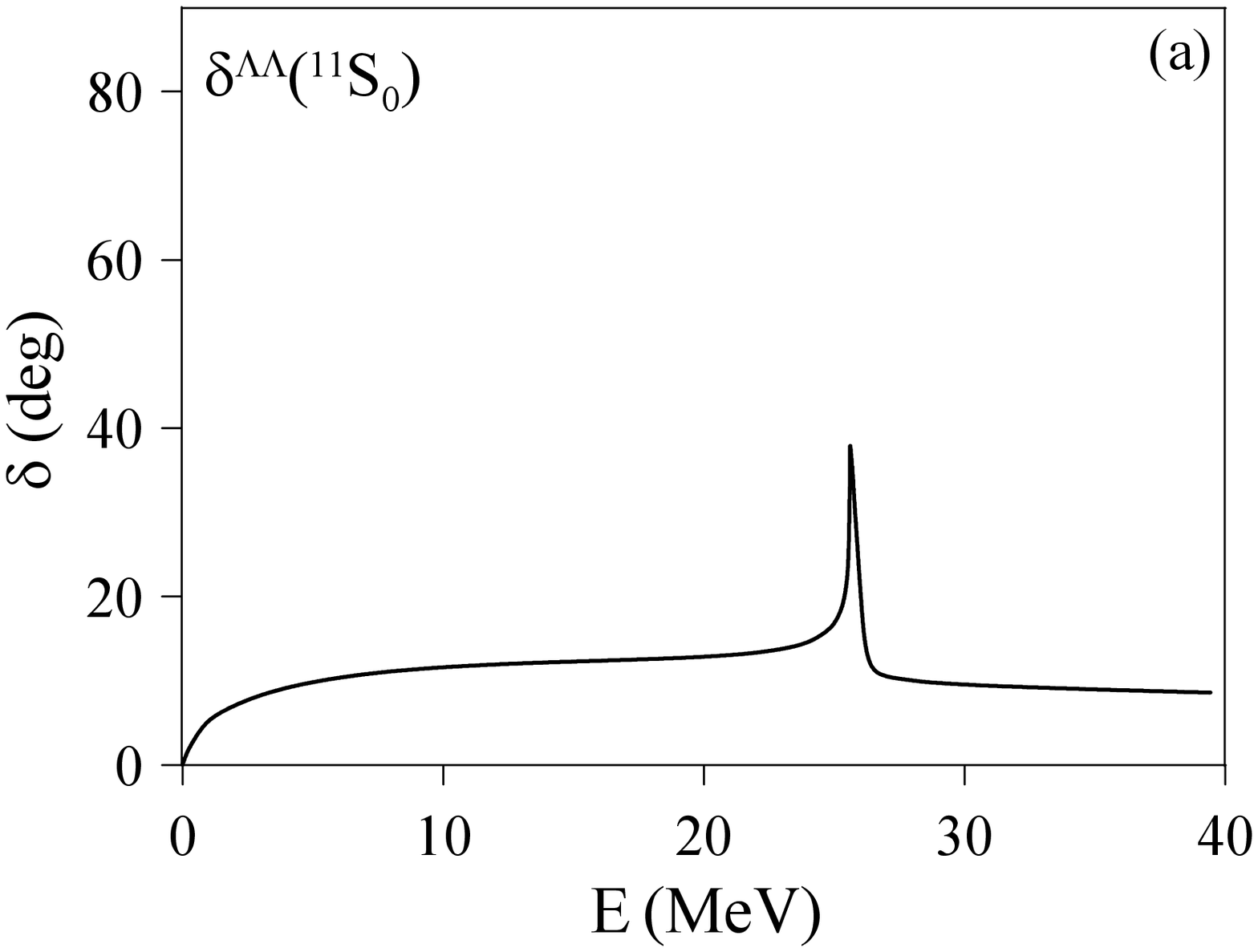}}
\resizebox{8.cm}{12.cm}{\includegraphics{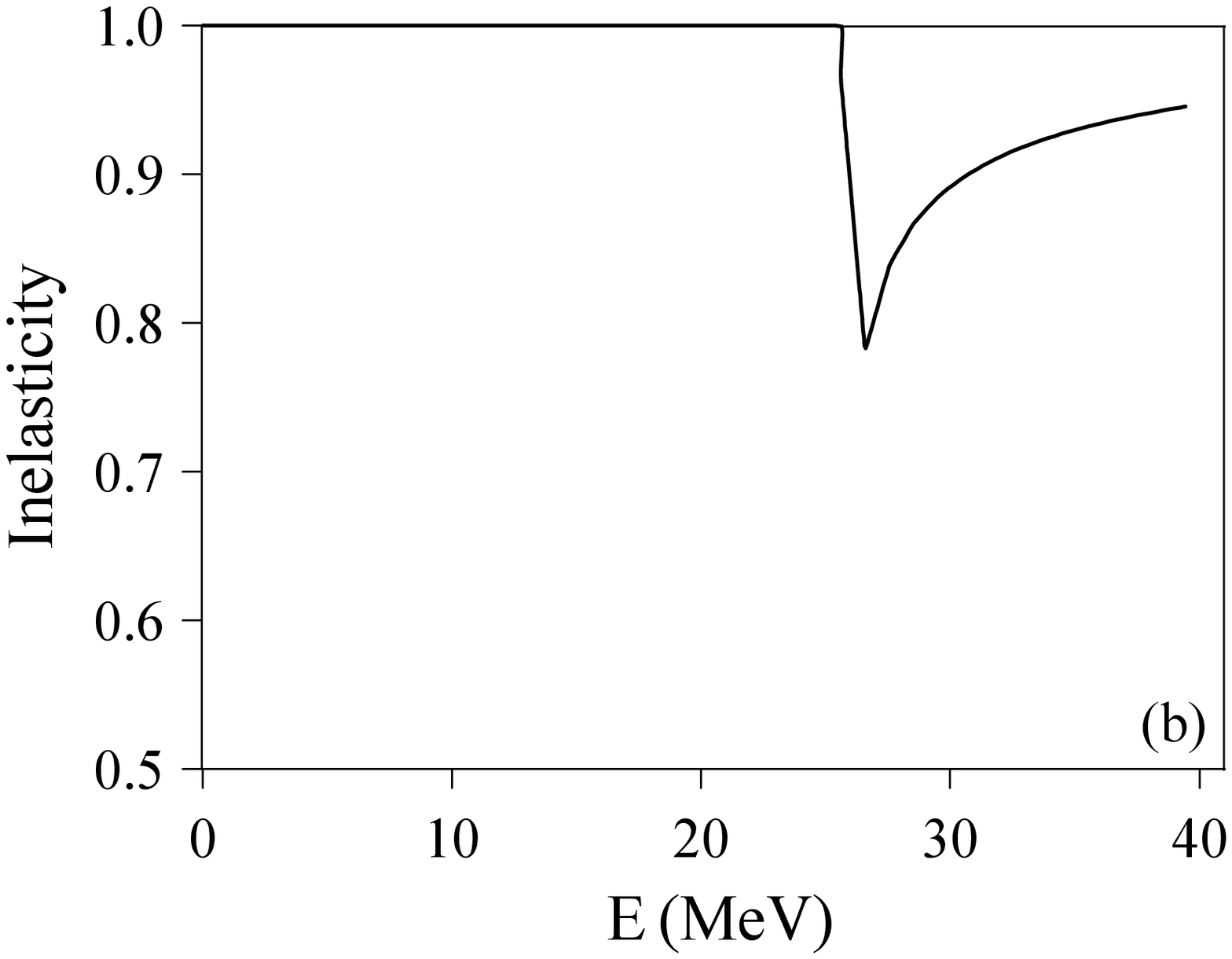}}\vspace*{-6.0cm}
\resizebox{8.cm}{12.cm}{\includegraphics{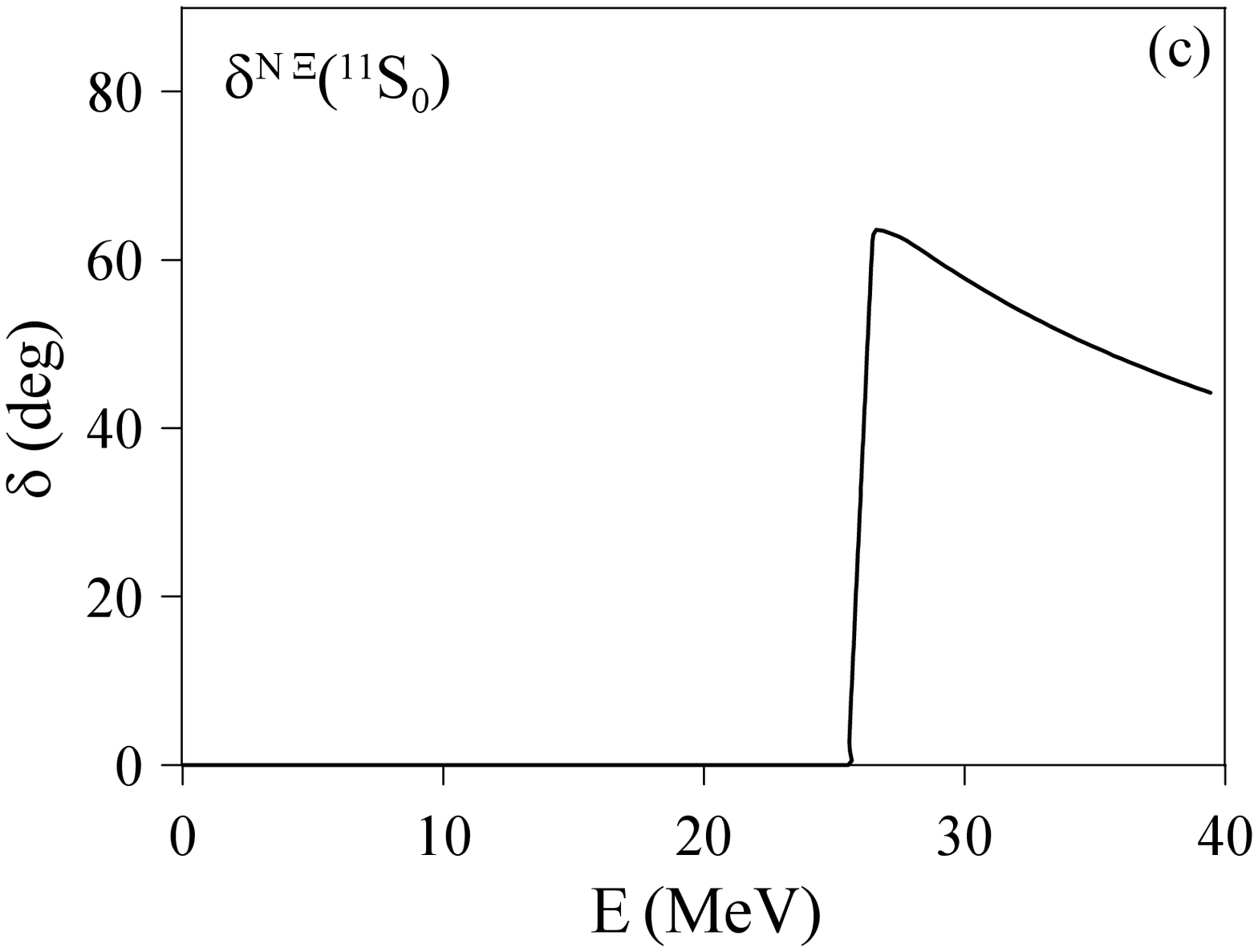}}
\vspace*{-6.0cm}
\caption{(a) $\Lambda\Lambda$ scattering phase shifts, (b) $\Lambda\Lambda$ inelasticity,
and (c) $N \Xi$ scattering phase shifts in the $(i,j^p)=(0,0^+)$ channel.}
\label{fig1}
\end{figure*}

In the case of the coupled $(0,0^+)$ $\Lambda\Lambda-\Xi N$ channel
we use a rank-two separable potential of the form~\cite{CAR1}
\begin{equation}
V_{ij}^{\rho\sigma} = g_i^\rho\rangle \lambda_{ij}\langle g_j^\sigma \, ,
\label{eq5}
\end{equation}
such that
\begin{equation}
t_{ij}^{\rho-\sigma} = g_1^\rho\rangle \tau_{ij}^{\rho-\sigma}\langle g_j^\sigma \, ,
\label{eq6}
\end{equation}
with
\begin{eqnarray}
\tau_{11}^{\Lambda\Lambda-\Lambda\Lambda} & = & 
\frac{-\lambda_{13}^2G^{\Xi N}
-\lambda_{11}(1-\lambda_{33}G^{\Xi N})}
{\lambda_{13}^2G^{\Lambda\Lambda}G^{\Xi N}
-(1-\lambda_{11}G^{\Lambda\Lambda})
(1-\lambda_{33}G^{\Xi N})} \nonumber \, , \\
\tau_{33}^{\Xi N - \Xi N} & = & 
\frac{-\lambda_{13}^2G^{\Lambda\Lambda}
-\lambda_{33}(1-\lambda_{11}G^{\Lambda\Lambda})}
{\lambda_{13}^2G^{\Lambda\Lambda}G^{\Xi N}
-(1-\lambda_{11}G^{\Lambda\Lambda})
(1-\lambda_{33}G^{\Xi N})} \, , \\
\tau_{13}^{\Lambda\Lambda-\Xi N} & = & \tau_{31}^{\Xi N-\Lambda\Lambda}=
\frac{-{\lambda_{13}}}
{\lambda_{13}^2G^{\Lambda\Lambda}G^{\Xi N}
-(1-\lambda_{11}G^{\Lambda\Lambda})
(1-\lambda_{33}G^{\Xi N})} \nonumber \, ,
\end{eqnarray}
\begin{table}[t]
\caption{Parameters $\alpha$ and $\beta$ (in fm$^{-1}$),
$\lambda_{11}$, $\lambda_{33}$, and $\lambda_{13}$
(in fm$^{-2}$) of the separable-potential model of the
coupled $(i,j^p)=(0,0^+)$ $\Lambda\Lambda -\Xi N$ two-body system.}
\begin{ruledtabular} 
\begin{tabular}{ccccccc} 
& $\alpha$  & $\beta$ & $\lambda_{11}$ & $\lambda_{33}$ & $\lambda_{13}$ &  \\
\hline
& 1.3465   & 1.1460  & $-$0.1390  & $-$0.3171  & 0.0977  &  \\
\end{tabular}
\end{ruledtabular}
\label{t2} 
\end{table}
\begin{table}[t]
\caption{Parameters $\alpha$ (in fm$^{-1}$)
and $\lambda$
(in fm$^{-2}$) of the separable-potential model of the
uncoupled $(i,j^p)$ 
$\Xi N$ two-body channels.}
\begin{ruledtabular} 
\begin{tabular}{ccccc} 
& Channel & $\alpha$  &  $\lambda$  &  \\
\hline
& $(0,1^+)$ & 1.41    & -0.117  &  \\
& $(1,0^+)$ & 7.333   & 22.97   &  \\
& $(1,1^+)$ & 0.803   & -0.016  &  \\
\end{tabular}
\end{ruledtabular}
\label{t3} 
\end{table}
and
\begin{eqnarray}
G^{\Lambda\Lambda} & = & 
\langle g_1^{\Lambda\Lambda}|G_0(1)|g_1^{\Lambda\Lambda}\rangle \nonumber \, , \\
G^{\Xi N} & = & 
\langle g_3^{\Xi N}|G_0(3)|g_3^{\Xi N}\rangle \,.
\end{eqnarray}
In this case we also use Yamaguchi-type form factors as
\begin{eqnarray}
g_1^{\Lambda\Lambda}(p) & = & \frac{1}{\alpha^2+p^2} \nonumber \, , \\
g_3^{\Xi N}(p) & = & \frac{1}{\beta^2+p^2} \, .
\end{eqnarray}
The parameters of the $\Lambda\Lambda - \Xi N$ model based on the latest
HAL QCD potentials are given in Table~\ref{t2}. In Figs.~\ref{fig1}(a),
(b), and (c)
we show the predictions for the $\Lambda\Lambda$ and $\Xi N$ phase
shifts as well as the inelasticity, which are rather similar to 
those of model $t/a=12$ of the HAL QCD Collaboration 
presented in Fig. 4 of Ref.~\cite{HALQCD}. The corresponding parameters
of the uncoupled $\Xi N$ models are given in Table~\ref{t3}.
Let us note that our results have been obtained by taking the nucleon mass as the
average of the proton and neutron masses and the $\Xi$ mass as the average 
of $\Xi^0$ and $\Xi^-$ masses. Thus, the $\Xi N$ and $\Xi NN$ thresholds are
25.6~MeV above the $\Lambda\Lambda$ and $\Lambda\Lambda N$ thresholds,
respectively. However, this threshold is 32 MeV for the HAL QCD
results~\cite{HALQCD}, since they use for the baryon masses the values
obtained from their lattice QCD study. Therefore, to compare our phase shifts 
with those of Ref.~\cite{HALQCD} one should keep in mind that 
the energy scale of Ref.~\cite{HALQCD} corresponds to those of Fig.~\ref{fig1}
multiplied by 1.25. The models of the $NN$ and $\Lambda N$
subsystems are the same of Ref.~\cite{GAR4}.

The coupled $\Lambda\Lambda N - \Xi NN$ three-body system presents
the special characteristic that each three-body 
component consists of two identical fermions
and a third one that is different. The homogeneous integral equations of this
system appropriate for the search of bound and resonant states
were derived in Ref.~\cite{GAR3} using a graphical method.
Using the new separable models of Tables I and II based in the HAL QCD
interactions we did not find any bound state below the
$\Xi d$ threshold, in agreement with the results of Ref.~\cite{HIYAM}. 
Therefore, we investigate the
possible existence of a resonance above the $\Xi d$ threshold by
calculating the $\Xi d$ scattering amplitude. 
 
We adopt the same convention as in Refs.~\cite{GAR3,GAR4}, i.e.,
particles $2$ and $3$ are identical and particle $1$ is the different
one in each three-body component. After the reduction for identical 
particles the inhomogeneous integral equations appropriate for $\Xi d$ 
elastic scattering take the following form
\begin{eqnarray}
\langle 1|T_1|\phi_0\rangle = &&
2\langle 1|t_1^{\Lambda\Lambda}|1\rangle 
\langle 1|3\rangle G_0(3)\langle 3|T_3|\phi_0\rangle+
\langle 1|t_{13}^{\Lambda\Lambda-N\Xi}|3\rangle 
\langle 3|1\rangle G_0(1) \langle 1|U_1|\phi_0\rangle
\nonumber \\ &&
-\langle 1|t_{13}^{\Lambda\Lambda-N\Xi}|3\rangle 
\langle 2|3\rangle G_0(3) \langle 3|U_3|\phi_0\rangle,
\nonumber \\
\langle 3|T_3|\phi_0\rangle = &&-\langle 3|t_3^{N\Lambda}|3\rangle 
\langle 2|3\rangle G_0(3)
\langle 3|T_3|\phi_0\rangle+\langle 3|t_3^{N\Lambda}|3\rangle
\langle 3|1\rangle G_0(1)
\langle 1|T_1|\phi_0\rangle,
\nonumber \\
\langle 1|U_1|\phi_0\rangle = && 2\langle 1|t_1^{NN}|1\rangle
\langle 1|3\rangle G_0(3)
\langle 3|U_3|\phi_0\rangle,
\nonumber \\ 
\langle 3|U_3|\phi_0\rangle = &&2\langle 3|t_3^{N\Xi}|\phi_0\rangle
\nonumber \\ &&
- \langle 3|t_3^{N\Xi}|3\rangle 
\langle 2|3\rangle G_0(3)
\langle 3|U_3|\phi_0\rangle
+\langle 3|t_3^{N\Xi}|3\rangle
\langle 3|1\rangle G_0(1)
\langle 1|U_1|\phi_0\rangle
\nonumber \\ &&
+2\langle 3|t_{31}^{N\Xi-\Lambda\Lambda}|1\rangle
\langle 1|3\rangle G_0(3)
\langle 3|T_3|\phi_0\rangle,
\label{eq91} 
\end{eqnarray}
where $|\phi_0\rangle$ is the initial state consisting of the
deuteron wave function times a $\Xi$ plane wave.

Using Eqs.~\eqref{eq3} and~\eqref{eq6} into the integral 
equations~\eqref{eq91} and 
introducing the transformations $\langle i|T_i|\phi_0\rangle =
\langle i|g_i^{\alpha_i}\rangle \langle i|X_i|\phi_0\rangle$ and
$\langle i|U_i|\phi_0\rangle =
\langle i|g_i^{\beta_i}\rangle \langle i|Y_i|\phi_0\rangle$ one
obtains the inhomogeneous one-dimensional integral equations
\begin{eqnarray}
\langle 1|X_1|\phi_0\rangle = 
&& 2\tau_1^{\Lambda\Lambda}\langle g_1^{\Lambda\Lambda}|1\rangle 
\langle 1|3\rangle G_0(3)\langle 3|g_3^{N\Lambda}\rangle
\langle 3|X_3|\phi_0\rangle
\nonumber \\ &&
+ \tau_{13}^{\Lambda\Lambda-N\Xi}\langle g_3^{N\Xi}|3\rangle 
\langle 3|1\rangle G_0(1) \langle 1|g_1^{NN}\rangle
\langle 1|Y_1|\phi_0\rangle
\nonumber \\ &&
- \tau_{13}^{\Lambda\Lambda-N\Xi}\langle g_3^{N\Xi}|3\rangle 
\langle 2|3\rangle G_0(3) \langle 3|g_3^{N\Xi}\rangle
\langle 3|Y_3|\phi_0\rangle,
\nonumber \\
\langle 3|X_3|\phi_0\rangle = && -\tau_3^{N\Lambda}
\langle g_3^{N\Lambda}|3\rangle\langle 2|3\rangle G_0(3)
\langle 3|g_3^{N\Lambda}\rangle
\langle 3|X_3|\phi_0\rangle
\nonumber \\ &&
+ \tau_3^{N\Lambda}\langle g_3^{N\Lambda}|3\rangle
\langle 3|1\rangle G_0(1)
\langle 1|g_1^{\Lambda\Lambda}\rangle
\langle 1|X_1|\phi_0\rangle,
\nonumber 
\end{eqnarray}
\begin{eqnarray}
\langle 1|Y_1|\phi_0\rangle = && 2\tau_1^{NN}\langle g_1^{NN}|1\rangle 
\langle 1|3\rangle G_0(3)
\langle 3|g_3^{N\Xi}\rangle
\langle 3|Y_3|\phi_0\rangle,
\nonumber \\ 
\langle 3|Y_3|\phi_0\rangle = && 2\tau_3^{N\Xi}\langle g_3^{N\Xi}|\phi_0\rangle 
\nonumber \\ &&
-\tau_3^{N\Xi}\langle g_3^{N\Xi}|3\rangle 
\langle 2|3\rangle G_0(3)
\langle 3|g_3^{N\Xi}\rangle
\langle 3|Y_3|\phi_0\rangle
\nonumber \\ &&
+ \tau_3^{N\Xi}\langle g_3^{N\Xi}|3\rangle
\langle 3|1\rangle G_0(1)
\langle 1|g_1^{NN}\rangle
\langle|Y_1|\phi_0\rangle
\nonumber \\ &&
+2\tau_{31}^{N\Xi-\Lambda\Lambda}\langle g_1^{\Lambda\Lambda}|1\rangle
\langle 1|3\rangle G_0(3)
\langle 3|g_3^{N\Lambda}\rangle\langle 3|X_3|\phi_0\rangle.
\label{eq92} 
\end{eqnarray}
If one neglects the inhomogeneous terms in Eqs.~\eqref{eq91} and~\eqref{eq92}
they become identical to Eqs.~(14) and~(15) of Ref.~\cite{GAR4}. The $\Xi d$
scattering amplitude normalized as in the Argand diagram is given by
\begin{equation}
F=-\pi q_0\nu\langle \phi_0|U_3|\phi_0\rangle,
\label{eq93}
\end{equation}
where $q_0$  and $\nu$ are, respectively, the $\Xi d$ on-shell momentum and
reduced mass. We solved the integral equations~\eqref{eq92} using the
standard method~\cite{AFNAN}, where the momentum variables are rotated
into the complex plane as $q_i\to q_i e^{-i\phi}$ and checked that the 
results are independent of the rotation angle $\phi$.
If the resonance lies below the $\Xi d$ threshold,
as it was the case in Ref.~\cite{GAR4}, the contour rotation
method allows to take simultaneously both the momentum
variables and the energy variable as complex so that one
can determine the position of the pole in the complex
plane. However, if the resonance lies above the $\Xi d$
threshold, like in the present case, the contour rotation
method works only if one takes the momentum variables
complex but leaves the energy variable real so that one can not
determine the position of the pole in the complex plane.
\begin{figure}[t]
\includegraphics[width=9.0cm]{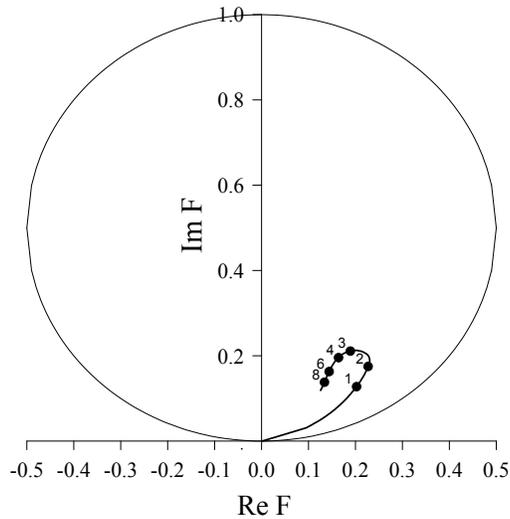}
\vspace*{-5.9cm} 
\caption{Argand diagram of the $\Xi d$ system between 0 and 10
MeV above the $\Xi d$ threshold. Some relevant energies, in MeV, are indicated.}
\label{fig2} 
\end{figure} 
\begin{figure}[b]
\vspace*{-3.9cm} 
\hspace*{4cm}\includegraphics[width=18.0cm]{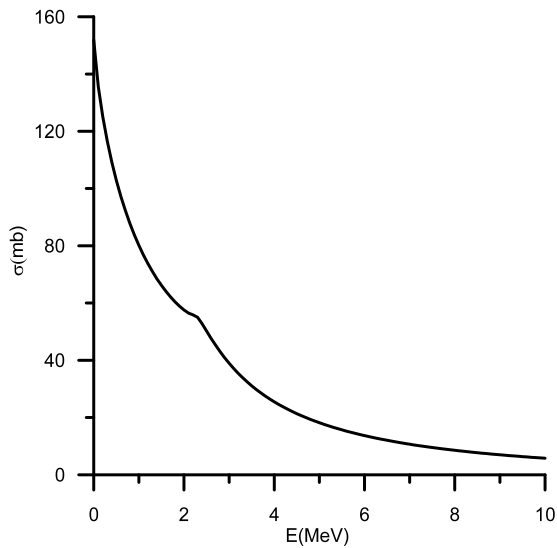}
\vspace*{-12.9cm} 
\caption{$\Xi d$ elastic cross section.}
\label{fig3}
\end{figure} 

We show in Fig.~\ref{fig2} the Argand diagram of the $\Xi d$ system between 0 and 10
MeV above the $\Xi d$ threshold where one sees the typical counterclockwise
behavior of a resonant amplitude. If one neglects the 
$(i,j^p)=(0,0^+)$ channel, the counterclockwise
behavior disappears which shows that the $H$-dibaryon channel
$(i,j^p)=(0,0^+)$ is basic for the existence of the three-body
$\Lambda\Lambda N - \Xi NN$ $S$ wave resonance.

As already mentioned in the introduction, the HAL QCD $\Xi N$ interactions have been 
recently used to study the possible existence of $\Xi NN$ bound 
states in Ref.~\cite{HIYAM} with negative results for the $(I,J^P)=(1/2,1/2^+)$
channel, in agreement with our findings in spite of using a different $NN$
interaction and a different method.

We have finally evaluated the $\Xi d$ elastic cross section 
as a function of energy where we have included not only 
the $(I,J^P)=(1/2,1/2^+)$ $\Xi d$ amplitude  
but also the $(1/2,3/2^+)$ amplitude, which is very small.
The result is shown in Fig.~\ref{fig3}.
As one can see, the resonance shows up as a change of slope 
of the cross section at an energy
around 2.3 MeV, i.e., close to  the $\Xi d$ breakup threshold
$\sqrt{S}=2m_N+m_\Xi$. The bump in the cross section would become
larger for a stronger $(i,j^p)=(0,0^+)$ transition potential,
and, as said above, it would disappear if the $(i,j^p)=(0,0^+)$ channel
is not considered or the two-body resonance in the $H$-dibaryon channel
would not exist. The $\Xi d$ cross section would allow
to discriminate among the different models for the strangeness $-$2
two-baryon interactions. It could be studied through the 
quasifree $\Xi^-$ production in the $(K^-,K^+)$ reaction
on a deuteron target~\cite{Tam01,Yam01}.

Let us finally note that if one drops the coupling to the 
$\Lambda\Lambda N$ channel Fig.~\ref{fig2} changes by about 
10 \% while keeping its shape rotating slightly to the right; 
similarly, in Fig.~\ref{fig3} the cross section
at $E=2.3$ MeV changes from 55 mb to 58 mb. 

It is interesting to compare this resonance 
with the nucleon-nucleon $^1D_2$ Hoshizaki resonance~\cite{HOSHI} 
which has a mass
close to $\sqrt{S}=m_N+m_\Delta$ since it arises
due to the process $NN\to\pi NN$
which is driven by the pion-nucleon $\Delta$ resonance~\cite{GA14}. 
The resonance we are studying here is driven by the 
$\Lambda\Lambda- \Xi N$  $H$-dibaryon resonance which appears either just below
or just above the $N\Xi$ threshold~\cite{HALQCD} so that it has a mass
$m_H=m_N+m_\Xi$.  Following the comparison with the Hoshizaki state
one expects that the
$\Lambda\Lambda N - \Xi NN$ resonance will have a mass close to
$\sqrt{S}=m_N +m_H=2m_N+m_\Xi$ which is
precisely the $\Xi d$ threshold in agreement with Figs.~\ref{fig2} and~\ref{fig3}.

In brief, we have shown that the possible existence of a 
$\Lambda\Lambda N - \Xi NN$ resonance would be highly sensitive 
to the $\Lambda\Lambda - \Xi N$ interaction. In particular,
by using a separable potential based on the most recent results 
of the HAL QCD Collaboration, characterized by the existence of a
resonance just below or above the $\Xi N$ threshold in the so-called 
$H$-dibaryon channel, $(i,j^p)=(0,0^+)$, 
a three-body resonance appears {2.3} MeV above the $\Xi d$ threshold.
A theoretical
and experimental effort to constrain the $\Lambda\Lambda - \Xi N$ interaction
is a basic ingredient to progress in our knowledge of the strangeness
$-$~2 sector.

\acknowledgments
This work has been partially funded by COFAA-IPN (M\'exico) and 
by Ministerio de Ciencia e Innovaci\'on
and EU FEDER under Contracts No. FPA2016-77177-C2-2-P, 
PID2019-105439GB-C22 and RED2018-102572-T.

\end{document}